\documentclass[pra,reprint,notitlepage,longbibliography]{revtex4-1}
\usepackage{amsmath}
\usepackage{amsthm}
\usepackage{graphicx}

\makeatletter

\providecommand{\tabularnewline}{\\}

\usepackage{hyperref}
\graphicspath{{.}{./figures/}}
\makeatother

\begin{document}
\global\long\def\sbn#1#2{\{#1\:|\:#2\}}%

\global\long\def\norm#1{\Vert#1\Vert}%

\global\long\def\abs#1{|#1|}%

\global\long\def\d{\mathrm{d}}%

\global\long\def\f#1#2#3{#1:\ #2\rightarrow#3}%

\global\long\def\k#1{\mathop{\mathrm{#1}}\nolimits}%

\global\long\def\braket#1#2{\Braket{#1|#2}}%

\global\long\def\bra#1{\Bra{#1}}%

\global\long\def\ket#1{\Ket{#1}}%

\global\long\def\u#1{\mathrm{\:#1}}%

\global\long\def\nd{\dot{N}}%

\title{Efficiency limits of electronically-coupled upconverter and \\ quantum ratchet solar cells using detailed balance}
\author{Emily Z. Zhang}
\altaffiliation{Current address: Department of Physics, University of Toronto, Toronto, ON, Canada}

\affiliation{Department of Physics, University of Ottawa, Ottawa, ON, Canada}
\author{Jacob J. Krich}
\affiliation{Department of Physics, University of Ottawa, Ottawa, ON, Canada}
\affiliation{School of Electrical Engineering and Computer Science, University
of Ottawa, Ottawa, ON, Canada}
\begin{abstract}
The intermediate band solar cell (IBSC) and quantum ratchet solar
cell (QRSC) have the potential to surpass the efficiency of standard
single-junction solar cells by allowing sub-gap photon absorption
through states deep inside the band gap. High efficiency IBSC and
QRSC devices have not yet been achieved, however, since introducing
mid-gap states also increases recombination, which can harm the device.
We consider the electronically coupled upconverter (ECUC) solar cell
and show that it can achieve the same efficiencies as the QRSC. Although
they are equivalent in the detailed balance limit, the ECUC is less
sensitive to nonradiative processes, which makes it a more practical
implementation for IB devices. We perform a case study of crystalline-silicon
based ECUC cells, focusing on hydrogenated amorphous silicon as the
upconverter material and highlighting potential dopants for the ECUC.
These results illustrate a new path for the development of IB-based
devices.
\end{abstract}
\maketitle

\section{Introduction}

\begin{figure}
\begin{centering}
\includegraphics[width=1\columnwidth]{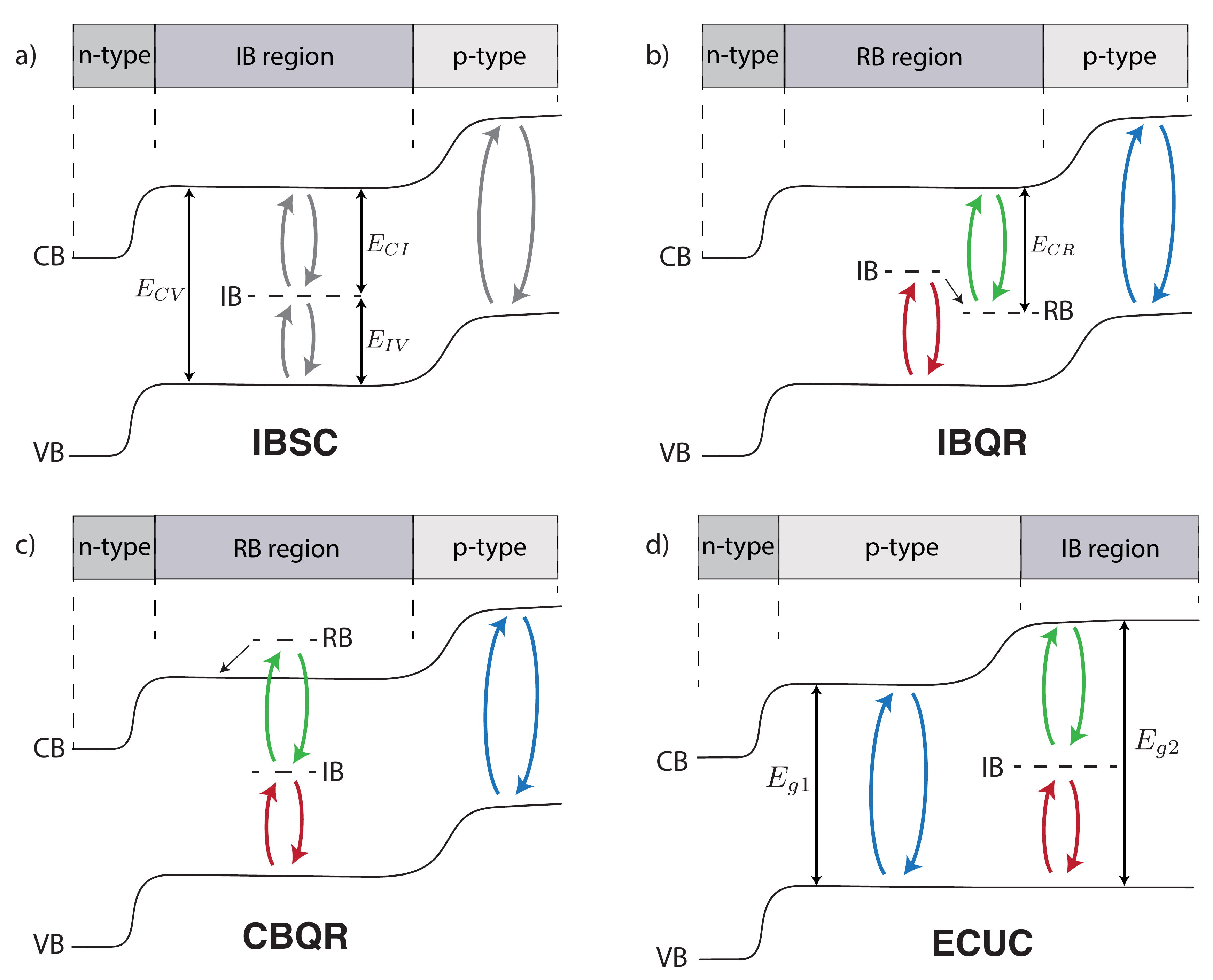}
\par\end{centering}
\caption{Band diagrams of (a) IBSC, (b) IBQR, (c) CBQR, and (d) ECUC. The red,
green, and blue processes for the ratchets and ECUC are equivalent
in detailed balance. \label{fig:bands}}
\end{figure}

Shockley and Queisser used the detailed balance (DB) formalism to
show that the efficiency of a solar cell made from a semiconductor
with a single band gap can never exceed 31\% under unconcentrated
black-body sunlight \citep{Shockley1961}. Intermediate band (IB)
materials -- semiconductors with allowed electronic states deep in
the gap, as shown in Figure \ref{fig:bands}a -- enable solar cells
to break this limit by absorbing sub-gap photons without harming the
voltage of the cell \citep{Luque1997}. In the radiative limit, the
maximum efficiency of an intermediate band solar cell (IBSC) at one
sun concentration is 47\%, significantly exceeding the Shockley-Queisser
limit \citep{Luque1997}. Several intermediate band devices have been
demonstrated, but high efficiencies have not been realized due to
nonradiative recombination \citep{Okada2015}.

The quantum ratchet (QR) solar cell has been proposed as an improved
implementation of the IBSC \citep{Yoshida2012}. The intermediate
band QR and conduction band QR implementations are shown in Figure
\ref{fig:bands}b-c, respectively. The original idea of a IBQR solar
cell is to increase the lifetime of the IB. In the case of the IBQR,
carriers relax from the IB to a ratchet band (RB), which can suppress
recombination to the valence band (VB). The ratchet also enables improved
voltage matching between the subgap transitions and the band-to-band
transitions \citep{Pusch2016,Pusch2019}. The CBQR has the ratchet
step above the conduction band edge, and an analogous valence band
QR (not shown) has the ratchet step below the valence band edge. All
three QR designs realize the voltage-matching improvements and can
achieve detailed balance maximum efficiencies of 48.5\% at one sun,
greater than that of IBSCs. There have, however, been few QR experimental
realizations and there are few suggestions for material systems \citep{Vaquero-Stainer2018}.

In both IBSC and QRSC devices, the IB and QR regions are added to
standard \emph{pn} junctions in hopes of increasing current in the
device, but if lifetimes are sufficiently short in the IB region,
the IBSC or QRSC may even have lower current than the reference \emph{pn}
junction. Both IBSCs and QRSCs have an \emph{n-}IB-\emph{p} architecture,
implying the holes created at the front of the cell must travel through
the IB region to be collected. If hole lifetimes in the IB or QR regions
are short, the nonradiative losses in the IB region will exceed the
extra current generation, making efficiencies less than for the \emph{pn-}diode
solar cell alone \citep{Krich2012,Wilkins2019}.

\begin{figure}
\centering{}\includegraphics[width=1\columnwidth]{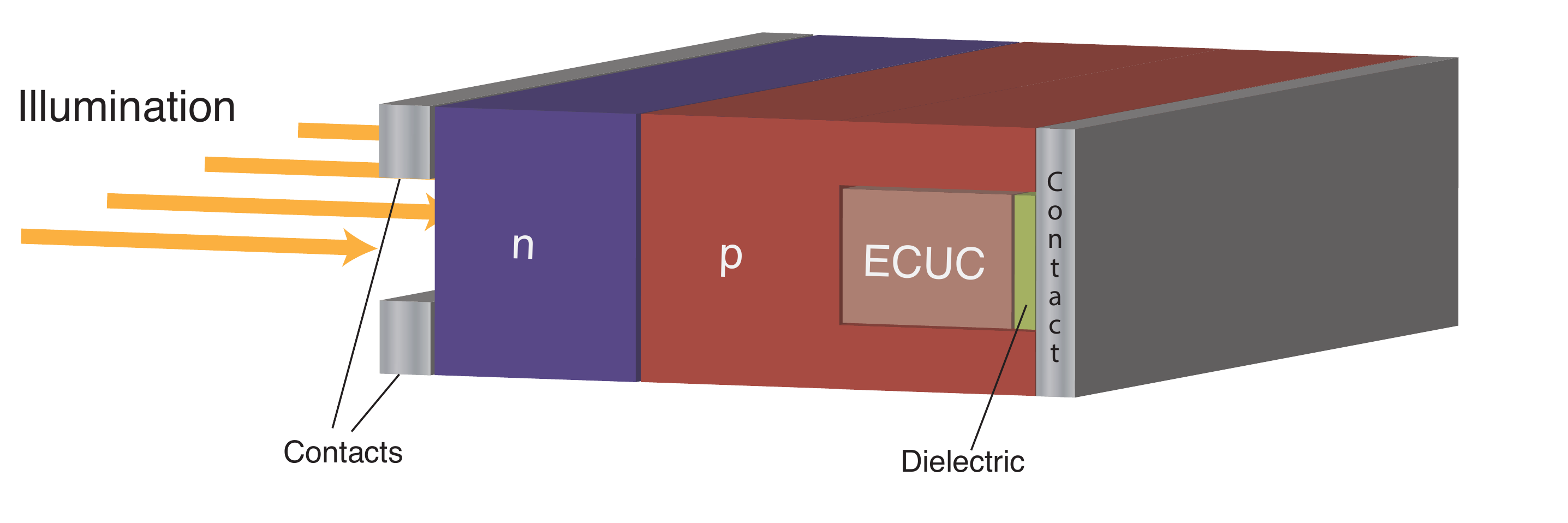}\caption{Schematic of a potential device architecture for the ECUC. \label{fig:Schematic_ECUC}}
\end{figure}

The electronically-coupled upconverter (ECUC) is a less-studied architecture,
which provides the potential to realize the same efficiency as a QRSC
while being less sensitive to nonradiative processes \citep{MacDonald2008,Harder2005}.
As shown in Figure \ref{fig:bands}d, the ECUC has an \emph{n-p-}IB
structure, with the IB region having a larger band gap than the standard
semiconductor, unlike in the IBSC and QRSC where the large band gap
$E_{CV}$ can be uniform through the device. As with IBSC and QRSC,
the ECUC allows absorption of subgap photons, with the resulting carriers
injected into the standard semiconductor. The minority carriers produced
by absorption in the \emph{pn} junction never transit the IB region,
so the current added from IB absorption can be obtained strictly as
an addition, and low quality upconverter material cannot harm the
cell as can occur in the IBSC/QRSC. However, the ECUC requires more
complicated 2D contacts to avoid extracting current from the IB, with
one possibility shown in Figure \ref{fig:Schematic_ECUC}. 

The detailed balance limiting efficiencies for the ECUC have not previously
been calculated. In this work, we demonstrate that the QRSC and ECUC
are mathematically equivalent in the DB limit, yet the ECUC may be
a more practical implementation in actual devices. We show that, as
with the QRSC, the ECUC configuration has the potential to exceed
IBSC efficiencies at 1 sun. We perform a global optimization showing
the maximum efficiencies possible as functions of $E_{g1}$ and $E_{g2}$
and also consider a case study of an ECUC based on crystalline silicon
(c-Si), the most widely used and studied PV material. We show that
there is potential to improve on c-Si solar cells using an ECUC.

\section{Detailed Balance Model}

We use the well-known detailed balance formalism to model the ECUC
and QRSC. We first show that in detailed balance, the ECUC and QRSC
are mathematically equivalent, then we use this method to compute
the limiting efficiencies for ECUC.

Detailed balance calculations assume all recombination is radiative,
carriers have infinite mobility, and the cell is thick enough to assure
full absorption of photons for each allowable transition. We further
assume perfect photon selectivity, with each photon absorbed only
by the highest-energy transition energetically permitted, to minimize
thermalization losses; this condition is called non-overlapping absorptions
and is not required for detailed balance \citep{Luque1997,Krishna2016,Cuadra2004}.
Since the carriers have infinite mobility,
\begin{equation}
\mu_{CV}=qV_{\text{ext}},\label{eq:mucv}
\end{equation}
where $q$ is the elementary charge, $\mu_{CV}$ is the quasi-Fermi
level difference between the electrons and holes, and $V_{\text{ext}}$
is the external voltage. We take $q=1$.

Another key assumption is that there is one electron-hole pair generated/lost
for each photon absorbed/emitted. Since all recombination events are
assumed to be radiative, this assumption allows the current in the
device to be written in terms of the photon fluxes $\phi$ in and
out of the device. These fluxes obey the modified Planck spectrum
\citep{Wurfel1982}
\begin{align}
\phi(E_{\min,AB},&E_{\max,AB},T,\mu_{AB})\\
&=\frac{2F}{h^{3}c^{2}}\int_{E_{\min,AB}}^{E_{\max,AB}}\frac{E^{2}dE}{e^{\left(E-\mu_{AB}\right)/kT}-1}, \nonumber
\end{align}
where the process between bands $A$ and $B$ absorbs photons with
energies between $E_{\min,AB}$ and $E_{\max,AB}$, $T$ is the temperature,
$\mu_{AB}$ is the chemical potential difference between carriers
in bands $A$ and $B$, $h$ is Planck's constant, $c$ is the speed
of light, $k$ is Boltzmann's constant, and $F$ is the geometrical
factor denoting the fraction of light incident on the cell. For the
sun, 
\begin{equation}
F_{\text{sun}}=X\cdot\pi\left(\frac{\text{radius of sun }}{\text{distance between earth and sun}}\right)^{2},
\end{equation}
where $X$ is the solar concentration factor, and for emission from
the cell, 
\begin{equation}
F_{\text{cell}}=\pi.
\end{equation}

In detailed balance, we have two photon sources: the sun and the cell.
We can denote the photons absorbed from the sun in transitions between
bands $A,B$ by 
\begin{equation}
\dot{N}_{AB}^{\text{sun}}=\phi\left(E_{\min,AB},E_{\max,AB},T_{s},0\right),
\end{equation}
and the photons emitted by the cell in transitions between bands $A,B$
by
\begin{equation}
\dot{N}_{AB}^{\text{cell}}=\phi\left(E_{\min,AB},E_{\max,AB},T_{a},\mu_{AB}\right),
\end{equation}
where $T_{s}$ is the solar radiation temperature, which we take to
be 6000~K and $T_{a}$ is the ambient temperature, which we take
to be 300~K. The current extracted from band $A$ is the difference
between absorbed and emitted photons involving band $A$,

\begin{equation}
J_{A}=\sum_{B}\pm\left(\dot{N}_{AB}^{\text{sun}}-\dot{N}_{AB}^{\text{cell}}(\mu_{AB})\right),
\end{equation}
 with the sign depending on whether the $AB$ absorption process creates
(+) or destroys (-) carriers in band $A$.

For all of the devices, the total current is the net current extracted
from either the CB or the VB, which are equal. For an ECUC, the total
current is 
\begin{equation}
J_{C}^{ECUC}=\nd_{CV}^{\text{sun }}-\nd_{CV}^{\text{cell}}(\mu_{CV})+\nd_{CI}^{\text{sun }}-\nd_{CI}^{\text{cell}}(\mu_{CI}).\label{eq:jc}
\end{equation}
We also assume that no current is extracted from the intermediate
band, so
\begin{equation}
J_{I}^{ECUC}=0=\nd_{IV}^{\text{sun }}-\nd_{IV}^{\text{cell}}(\mu_{IV})-\nd_{CI}^{\text{sun }}+\nd_{CI}^{\text{cell}}(\mu_{CI}).\label{eq:ji}
\end{equation}
Note that the CI processes in Eq.~\ref{eq:ji} enter with the negative
sign, as optical absorption from IB to CB removes an IB carrier. With
equations \ref{eq:mucv},\ref{eq:jc},\ref{eq:ji}, and the fact that
\begin{equation}
\mu_{CV}=\mu_{CI}+\mu_{IV},
\end{equation}
we can solve for the chemical potentials and compute $J(V)$. These
equations are of the same form as in the original IBSC calculation
\citep{Luque1997}, but the ECUC has different band gaps in the different
regions. Note that the $\mu_{CV}$ terms use $E_{g1}$ as their lower
threshold. 

For an IBQR, we assume the carriers in the IB and RB share a common
quasi-Fermi level, so $\mu_{CI}=\mu_{CR}$ \citep{Yoshida2012}. Then,
the net current from the CB is

\begin{equation}
J_{C}^{IBQR}=\nd_{CV}^{\text{sun }}-\nd_{CV}^{\text{cell}}(\mu_{CV})+\nd_{CR}^{\text{sun }}-\nd_{CR}^{\text{cell}}(\mu_{CR}),\label{eq:jcqr}
\end{equation}
and the net current in the IB is

\begin{equation}
J_{I}^{IBQR}=0=\nd_{IV}^{\text{sun }}-\nd_{IV}^{\text{cell}}(\mu_{CR})-\nd_{CR}^{\text{sun }}+\nd_{CR}^{\text{cell}}(\mu_{CR}).\label{eq:jiqr}
\end{equation}

These equations for the ECUC and IBQR are equivalent. As shown in
Figure \ref{fig:bands}d, $E_{CI}+E_{IV}=E_{g2}$ for the ECUC. If
we choose $E_{CV}$ for the IBQR to equal $E_{g1}$ for the ECUC then
the first two terms in each of Eqs.~\ref{eq:jcqr} and \ref{eq:jiqr}
are equal to the equivalent terms in Eqs.~\ref{eq:jc} and \ref{eq:ji}.
Further, if $E_{CR}$ for the IBQR equals $E_{CI}$ for the ECUC,
and $E_{IV}+E_{CR}$ for the IBQR equals $E_{g2}$ for the ECUC, then
the last two terms in each of those equations become equivalent. Therefore
the ECUC equations are equal to the IBQR equations. Similarly, if
$E_{IV}+E_{RI}=E_{g2}$ for the CBQR or $E_{IR}+E_{CI}=E_{g2}$ for
the valence band QR (VBQR), then the equations also become equivalent
to the ECUC. Since the equations for QR and ECUC are no different
in detailed balance, the limiting efficiencies are also the same.

\begin{figure}
\begin{centering}
\includegraphics[width=1\columnwidth]{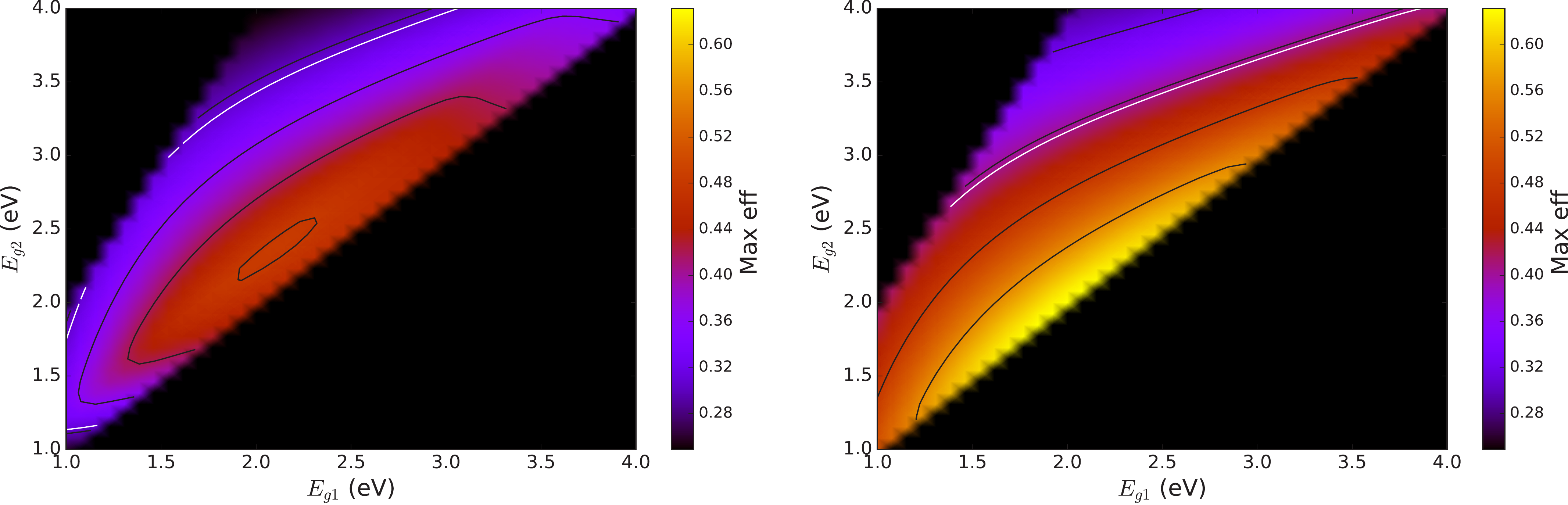}
\par\end{centering}
\caption{Maximum ECUC efficiency in detailed balance, with optimized $E_{I}$,
at 1 sun concentration (left) and at full concentration ($X=46200$)
(right). The Shockley-Queisser limits of 31\% ($X=1$) and 40.7\%
($X=46200$) are shown with the white contours. Note that an ECUC
can only be beneficial if $E_{g1}\protect\leq E_{g2}\protect\leq2E_{g1}$,
with the second inequality from the requirement that both sub-gap
transitions have energy thresholds below $E_{g1}$. \label{fig:Maximum-ECUC-efficiency}}
\end{figure}

Figure \ref{fig:Maximum-ECUC-efficiency} shows the maximum ECUC efficiencies
at $X=1$ and $X=1/F_{\text{sun}}=46200$, which is the maximum value.
The peak efficiencies and band gaps for these cases are shown in Table
\ref{tab:global1sun}. The diagonal border at $E_{g1}=E_{g2}$ represents
standard IB solar cells, and at one sun concentration (left), the
detailed balance efficiency is highest at $E_{g1}\ne E_{g2}$. This
result indicates that the ECUC has higher limiting efficiency than
IBSC, similar to QR \citep{Yoshida2012}, spectrally-selective reflectors
\citep{Strandberg2010}, and overlapping absorptions \citep{Krishna2016}.
Therefore, the ECUC can exceed both the IBSC limit and the Shockley-Queisser
limit. Figure \ref{fig:Maximum-ECUC-efficiency} shows that there
is a wide range of band gaps that can potentially achieve this goal.

At full concentration, the highest efficiency lies on $E_{g1}=E_{g2}$,
so there is no gain from ECUC compared to a standard IBSC architecture.
Both the ECUC and the IBSC significantly exceed the single junction
efficiency limit, which has motivated interest in combining IBSC with
concentrator systems \citep{Luque2010,Sogabe2014}.

\begin{table}
\caption{\label{tab:global1sun}Maximum efficiencies
with a blackbody spectrum at 1 sun and full concentration. Note that
there is a symmetry for $E_{I}$ mirrored below and above $E_{g2}/2$;
we take the upper values for $E_{I}$. }

\centering{}%
\begin{tabular}{cccccc}
\hline 
System & $X$ & $E_{g1}$ (eV) & $E_{g2}$ (eV) & $E_{I}$ (eV) & Efficiency\tabularnewline
\hline 
Single-junction & 1 & 1.31 & - & - & 31.0\%\tabularnewline
IB solar cell & 1 & 2.42 & - & 1.49 & 46.8\%\tabularnewline
ECUC solar cell & 1 & 2.08 & 2.36 & 1.42 & 48.5\%\tabularnewline
Single-junction & 46200 & 1.11 & - & - & 40.7\%\tabularnewline
IB solar cell & 46200 & 1.95 & - & 1.24 & 63.2\%\tabularnewline
ECUC solar cell & 46200 & 1.95 & 1.95 & 1.24 & 63.2\%\tabularnewline
\hline 
\end{tabular}
\end{table}

\section{Case study: ECUC using c-Si}

In this section, we perform a case study of a potential ECUC using
silicon as the front $pn$-diode material, since c-Si is an extremely
well-understood material. Adding only an intermediate band to an \emph{n-}IB-\emph{p}
c-Si solar cell actually harms the efficiency of the cell, even in
the detailed balance limit \citep{Krishna2016}. That failure occurs
because of silicon's small band gap and the assumption of non-overlapping
absorptions. Figure \ref{fig:Maximum-ECUC-efficiency}, however, shows
that even with $E_{g1}$ equal to the band gap of c-Si, 1.12 eV, the
ECUC allows considerable improvement over the Shockley-Queisser limit.
First, we study the optimal range for $E_{g2}$ for an ECUC on silicon.
Second, we consider an ECUC made of hydrogenated amorphous silicon
(a-Si), which is a higher band-gap material frequently used for heterojunctions
with c-Si. We perform a search for the best-suited $E_{I}$ for an
a-Si upconverter on c-Si.

\begin{figure}
\begin{centering}
\includegraphics[width=1\columnwidth]{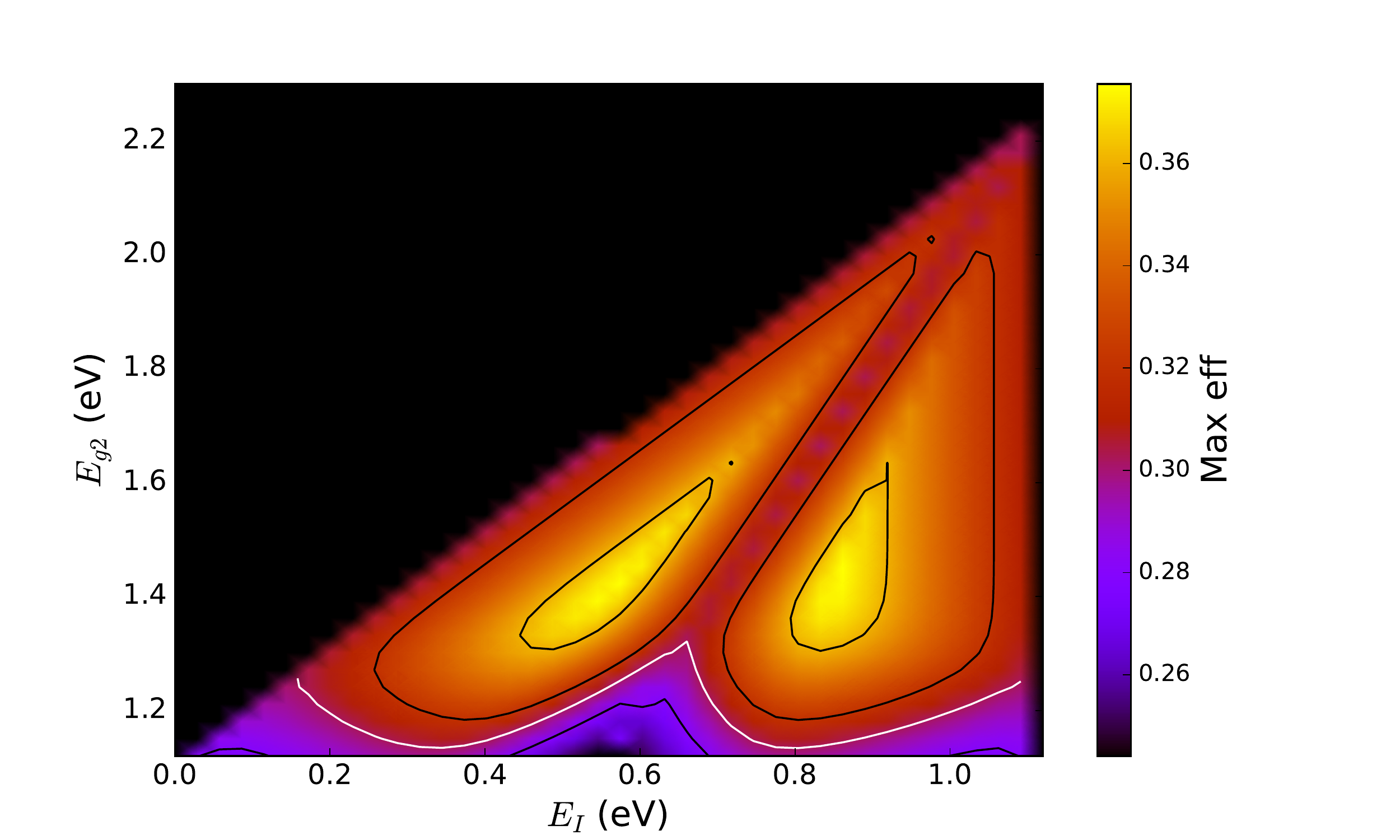}
\par\end{centering}
\caption{Maximum ECUC efficiency in detailed balance, with $E_{g1}=1.12\protect\u{eV}$
as a function of $E_{g2}$ and $E_{I}$ at 1 sun concentration. The
detailed balance efficiency limit for $E_{g}=1.12$~eV is shown with
the white contour.\label{fig:Maximum-ECUC-efficiency-1} Note
that the data cutoff at the diagonal (black) occurs because the ECUC
requires $E_{I}>E_{g2}-E_{g1}$ and $E_{I}<E_{g1}$.}
\end{figure}

Figure \ref{fig:Maximum-ECUC-efficiency-1} shows the maximum ECUC
efficiency with $E_{g1}=1.12$~eV as a function of $E_{g2}$ and
$E_{I}$. The peak efficiencies and band gaps are shown in Table \ref{tab:Maximum-efficiencies-for}.
The optimal range of $E_{g2}$ lies approximately between 1.3 and
1.6~eV, with the maximum efficiency at $E_{g2}=1.47\u{eV}$, with
$E_{I}$ near $0.9\u{eV}$. As $E_{g2}$ approaches $E_{g1}$, we
recover the IBSC efficiency, which is lower than the Shockley-Queisser
limit for a device with $E_{g}=1.12$~eV. Note that when $E_{g2}>1.3\u{eV}$,
the ECUC improves efficiencies for all values of $E_{I}$. For a large
range of band gaps, it is possible to significantly exceed the SQ
limit; therefore, there is potential for high efficiency silicon devices
if an ECUC is added.

\begin{table}
\caption{\label{tab:Maximum-efficiencies-for}Maximum detailed balance efficiencies
for c-Si ($E_{g}=1.12\protect\u{eV}$) at 1 sun concentration. Note
that there is a symmetry for $E_{I}$ mirrored below and above $E_{g2}/2$;
we take the upper values for $E_{I}$.}

\centering{}%
\begin{tabular}{ccccc}
\hline 
System & $E_{g1}$ (eV) & $E_{g2}$ (eV) & $E_{I}$ (eV) & Efficiency\tabularnewline
\hline 
Single-junction & 1.12 & - & - & 30.2\%\tabularnewline
IB solar cell & 1.12 & - & 0.85 & 29.7\%\tabularnewline
ECUC solar cell & 1.12 & 1.47 & 0.86 & 37.4\%\tabularnewline
\hline 
\end{tabular}
\end{table}

\begin{figure}
\begin{centering}
\includegraphics[width=1\columnwidth]{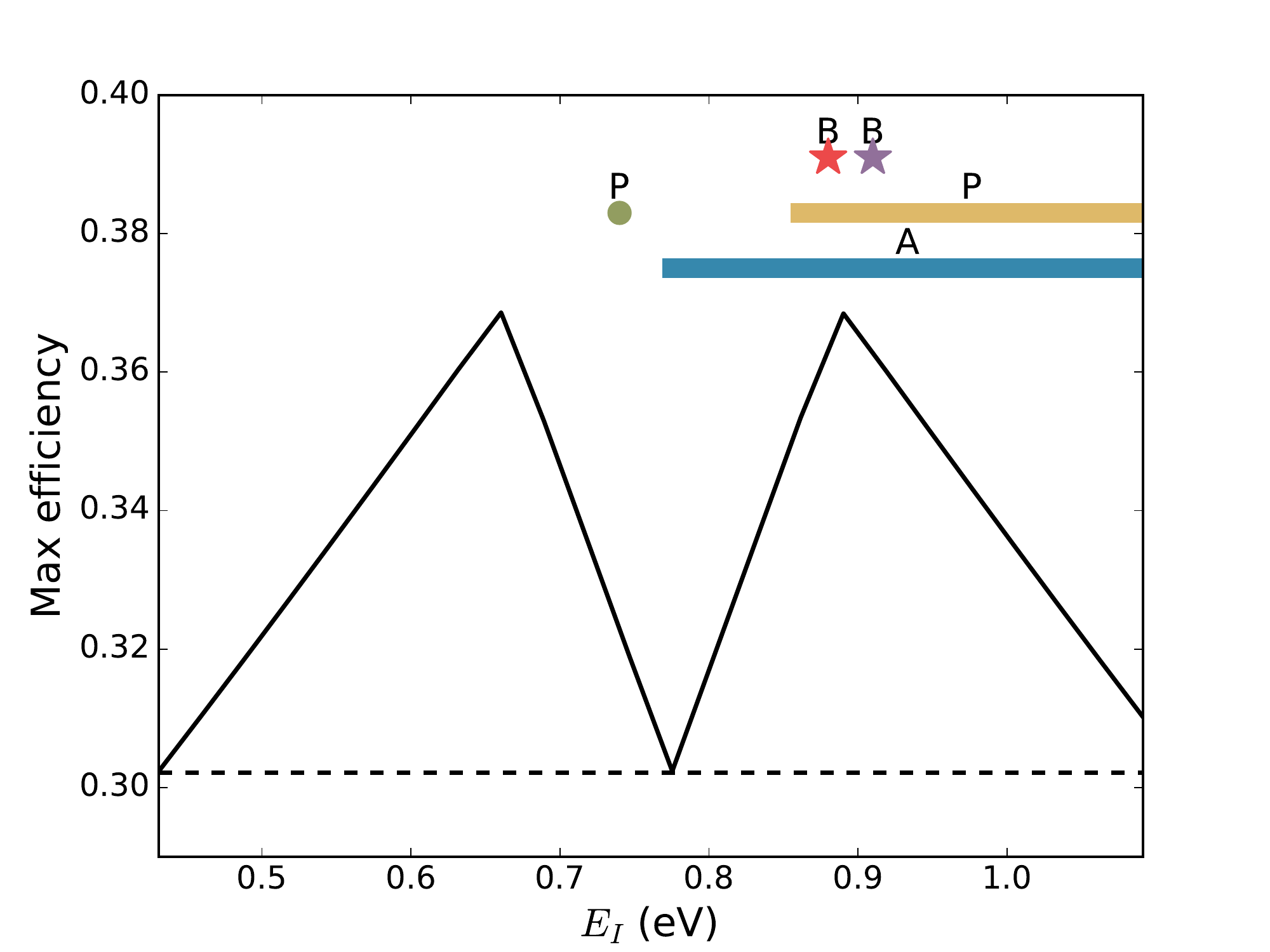}
\par\end{centering}
\caption{Maximum ECUC efficiency vs.~$E_{I}$ for c-Si ($E_{g1}=1.12\protect\u{eV}$)
and an a-Si upconverter ($E_{g2}=1.55\protect\u{eV}$). The black
dashed line shows the single-junction detailed balance efficiency
with $E_{g}=1.12$~eV. The potential dopants are labelled at their
respective $E_{I}$. Doping with P is shown with the green dot (optical
\citep{Street1984}) and a range of values with the yellow line (electrical
activation \citep{Matsuda1980}). Thermal activation energies for
B are shown with stars, with red corresponding to doping with BF$_{3}$
\citep{Mahan1983} and purple to B$_{2}$H$_{6}$ \citep{Street1984}.
The blue line shows the range of $E_{I}$ from thermal activation
for alkali dopants, including Na, K, Rb, and Cs \citep{LeComber1980}.
\label{fig:Maximum-ECUC-efficiency-2}}
\end{figure}

A promising upconverter material is amorphous silicon, since its band
gap of $E_{g2}=1.55\u{eV}$ falls in the high-efficiency range \citep{Carlson1976}
, and a-Si on c-Si devices are routinely made \citep{Mishima2011}.
Figure \ref{fig:Maximum-ECUC-efficiency-2} shows DB efficiency as
a function of $E_{I}$ of a device using c-Si and an a-Si ECUC. All
values of $E_{I}$ between $E_{g2}-E_{g1}=0.43$~eV and $E_{g1}=$1.12~eV
give improved efficiencies over the bare c-Si cell. Doping of a-Si
is more complicated than in crystalline semiconductors, as dopants
can induce local coordination changes and dangling bonds, and the
structures vary depending on deposition method \citep{Carlson1990}.
The resulting $E_{I}$ for a dopant in a-Si can thus vary considerably
depending on a-Si deposition and dopant precursor and pressure \citep{Carlson1990}.
This variation could allow tuning of ECUC energy levels, which is
not generally possible in crystalline semiconductor:dopant materials.
To date, devices based on doped a-Si have generally desired shallow
dopants, as in c-Si, so the most-studied dopants are those that produce
relatively shallow states in the band gap, to give high conductivities.
For an ECUC, optically active midgap states are desirable, which is
the opposite of the standard case.

Figure \ref{fig:Maximum-ECUC-efficiency-2} also shows estimated energetic
positions for some common dopants in a-Si. The most studied dopants
include boron and phosphorus as acceptors and donors, respectively,
as in c-Si. Even when a-Si has tetrahedrally coordinated silicon,
the bond angle distortions tend to make dopant energy levels lie deeper
in the gap than in c-Si \citep{Nichols1987}. As an acceptor, B doping
using B$_{2}$H$_{6}$ or BF$_{3}$ gives an electrical activation
energy of $E_{I}=0.88-0.91\u{eV}$, with a higher concentration of
active dopant states formed from the BF$_{3}$ precursor \citep{Mahan1983,Street1984}.
As a donor, P doping using PH$_{3}$ gives optical absorption in a
band around $E_{g2}-E_{I}=0.81\u{eV}$ \citep{Street1984}. As can
be seen in Fig.~\ref{fig:Maximum-ECUC-efficiency-2}, this energy
level appears close to the middle of the band gap, which allows only
minimal improvement in these detailed balance calculations. That dip
in efficiency for $E_{I}\approx E_{g2}/2$ is an artifact of the non-overlapping
absorption condition, as one of the subgap transitions becomes artificially
depleted of photons when $E_{I}$ is close to mid-gap. Removing the
non-overlapping absorption requirement, which is only a simplifcation
for theoretical analysis, reduces the penalty for IB's at mid-gap
\citep{Cuadra2004,Krishna2016}, so this mid-gap $E_{I}$ can still
be beneficial for the ECUC. Doping with P has also been shown to produce
thermal activation energies ranging from 0.74~eV to 0.27~eV, depending
on concentration of the precursor, with higher activation energies
at lower doping concentrations \citep{Matsuda1980}. Alkali atoms
as donors, including Na, K, Rb, and Cs, have been shown to produce
thermal activation energies that are similar to each other, ranging
from $0.80\u{eV}$ to $0.20\u{eV}$, again with higher activation
energies at lower dopant concentrations \citep{LeComber1980}. We
interpret these activation energies to be $E_{g2}-E_{I}$. These values
contain overlap with the optimal efficiency range for a c-Si/a-Si
ECUC. A working ECUC must be optically thick for the subgap photons,
which requires either a high dopant concentration or a thick absorber
layer. If high dopant concentration is required, the alkali dopant
energy levels may be less than than $E_{g2}-E_{g1}$ and thus outside
of the useful energy range. 

The combination of c-Si and a-Si has great potential to make a working
ECUC that can improve the efficiency of c-Si solar cells. To realize
this potential, the energetic position of those defect states and
their optical properties must be characterized, both for the common
electrical dopants and possibly a much larger range of potential IB-forming
dopants. A wide array of elements may be interesting for a-Si based
ECUC, just as a wide array of dopants may be useful for c-Si based
IBSC's \citep{Sullivan2015}.

\section{Conclusions}

The ECUC has the potential to improve IB solar cell designs. Its maximum
detailed balance efficiency is equal to that of a QRSC, and it may
be easier to produce. Though DB calculations do not consider non-radiative
processes, they give upper bounds on the efficiency of all photovoltaic
devices. At low solar concentration, ECUC has a higher limiting efficiency
than IBSC. This effect is realized in the c-Si case at one sun, where
an IBSC with non-overlapping absorptions cannot improve on a standard
single-gap solar cell, but an ECUC permits significantly improved
efficiency. At high concentration, the DB efficiency limits of IBSC,
ECUC, and QRSC are all the same, with a significant gain compared
to a single junction device. Moving beyond DB, the ECUC architecture
allows improved efficiency even with materials having significant
nonradiative recombination. It is thus a promising architecture to
pursue for near-term development of IB-based devices. The case of
a-Si on c-Si provide a promising platform for developing an ECUC with
the potential to significantly improve silicon-based solar cell efficiencies.
\begin{acknowledgments}
We acknowledge helpful conversations with Daniel MacDonald and Wenjie
Yang and support from the Natural Sciences and Engineering Research
Council of Canada.
\end{acknowledgments}

\bibliography{ECUC}

\end{document}